\documentclass[twocolumn,prX,showpacs]{revtex4}

\usepackage{amssymb,amsmath}
\usepackage{openbib}
\usepackage{amsthm}
\usepackage{epsfig}
\usepackage{subeqnarray}
\usepackage{graphicx}

\usepackage{notoccite} 
\usepackage{hyperref}

\newcommand{\ket}[1]{{|#1\rangle}}
\newcommand{\bra}[1]{{\langle#1|}}

\DeclareMathOperator{\Tr}{Tr}
\newcommand{\eq}[1]{{eq.(\ref{#1})}}
\renewcommand{\theequation}{{\arabic{section}.\arabic{equation}}}  

\begin{document}

\title{Measures of Quantum State Purity and Classical Degree of Polarization}

\author{Omar Gamel}\email{ogamel@physics.utoronto.ca}\author{Daniel F. V. James} 
\address{Department of Physics, University of Toronto, 60 St. George St., Toronto, Ontario, Canada M5S 1A7.}

\revised{October 6, 2013}

\begin{abstract}
There is a well known mathematical similarity between two dimensional classical polarization optics and two level quantum systems, where the Poincare and Bloch spheres are identical mathematical structures. This analogy implies classical degree of polarization and quantum purity are in fact the same quantity. We make extensive use of this analogy to analyze various measures of polarization for higher dimensions proposed in the literature, and in particular, the $N=3$ case, illustrating interesting relationships that emerge as well the advantages of each measure. We also propose a possible new class of measures of entanglement based on purity of subsystems.
\end{abstract}


\maketitle

\renewcommand{\theequation}{\arabic{equation}}

\section{Introduction}
There exists a well known mathematical similarity between classical polarization optics in two dimensions and quantum two level systems. The Stokes vector and Poincar\'{e} sphere on the one hand \cite{stokes, poincare} are analogous to the Bloch vector and Bloch sphere on the other \cite{bloch}. The Pancharatnam phase in classical optics systems \cite{pancha} corresponds to the Berry phase in quantum systems \cite{berry}. 

The point at the origin of the Poincar\'{e} sphere represents a completely unpolarized beam of light, while any point on the surface of the sphere represents a completely polarized beam. In the case of the Bloch sphere, the origin represents the maximally mixed state while any point on the surface represents a pure state. This suggests an additional analogy between the two systems, that classical polarization is analogous to quantum purity, and measures of the two quantities should therefore be identical. This analogy between polarization and purity has been discussed by some authors \cite{aiello}. Some authors have suggested the use of Bell's measure, commonly used in tests of quantum non-locality, to quantify classical optical coherence (polarization) \cite{saleh}. Others have suggested a particular measure of polarization in higher dimensions \cite{friberg08}. However despite this analogy, the most widely used measurements of classical degree of polarization and quantum mechanical purity are different.

In this paper, we start in section \ref{s2:polarizationpurity} by discussing in more detail the analogy between the classical and quantum cases, demonstrating that classical degree of polarization and quantum mechanical purity should be identical quantities. Section \ref{s3:measures} introduces many existing measures of purity and polarization in generic $N$ dimensions, with particular attention to the three dimensional case. In doing so, we come across measures for quantum purity from quantum mechanics, namely the standard purity and von Neumann purity \cite{nielsenchuang}. We also analyze measures of polarization in three dimensions due to Barakat \cite{barakat83}, Friberg et. al. \cite{friberg02}, and Wolf et al. \cite{wolf04}. 

We then proceed in section \ref{s4:comparison} to compare these measures analytically and numerically, giving them physical interpretations where possible. Our analysis adds to and clarifies much of the discussion on measures of higher dimensional polarization in the literature \cite{bergman, wolf04b, bjork05, luis05, friberg09, friberg10, sheppard, gil10}, classifying the measures and analyzing their relationship. 

In section \ref{s5:entanglement}, we point out that the entanglement of a bipartite pure state can be thought of as the purity of a subsystem once the other subsystem is traced out. This suggests that using unconventional measures of purity, we can create new and interesting measures of entanglement.

\section{Polarization of beams and purity of qubits}\label{s2:polarizationpurity}

\subsection{Classical polarization states}

Consider a classical beam of light propagating in the $z$ direction. The complex electric field values in the $x$ and $y$ direction are taken to be probabilistic ensembles given by complex analytic signals $E_1(\bold{r},t)$ and $E_2(\bold{r},t)$ respectively, where $\bold{r}$ is the position vector.

The polarization state of the beam of light is given by the 2$\times$2 polarization matrix $\Phi(\bold{r},t)$, defined as
\begin{equation}
\Phi_{ij}=\langle E_iE_j^* \rangle, \hspace{20pt} i=1,2.
\label{phidef}
\end{equation} 
where position and time dependence have been suppressed. If one thinks of $E_1$ and $E_2$ as random variables, then $\Phi$ is their variance-covariance matrix. 

Alternatively, the four element Stokes vector $S$ can be used to represent the polarization state \cite{stokes}. It is related to $\Phi$ by
\begin{align}
S_\mu &=Tr[\Phi \sigma^\mu ]. \label{stokeseq}
\end{align} 
where $\sigma^0$ is the identity matrix, and $\sigma^1$, $\sigma^2$, and $\sigma^3$ are the three Pauli matrices $\sigma^z$, $\sigma^x$, and $\sigma^y$ respectively. Einstein summation notation has been used, i.e. repeated indices are summed over. Lowercase Latin letters run from 1 to 2 (corresponding to the two Cartesian components of the transverse field), while lowercase Greek letters run from 0 to 3. The polarization matrix or Stokes vector contain all the physical information about the polarization state of the beam \cite{brosseau98}, and are different ways of mathematically representing the same information.

The inverse relationship to \eq{stokeseq} is given by
\begin{align}
\Phi &= \frac{1}{2}S_\mu\sigma^\mu \nonumber\\
&=\frac{1}{2}\begin{bmatrix} S_0 + S_1 & S_2 - iS_3 \\ S_2 + iS_3 & S_0 - S_1\end{bmatrix}.
\label{phistokeseq}
\end{align}

Equation (\ref{phidef}) implies that $\Phi$ is a positive and Hermitian matrix. The positivity of $\Phi$ implies $\det(\Phi) \ge 0$, which when applied to \eq{phistokeseq} implies the following condition on the Stokes parameters \cite{bornwolf}:
\begin{align}
S_1^2 + S_2^2 + S_3^2 \le S_0^2.
\label{poincaresphere}
\end{align}
The inequality in \eq{poincaresphere} shows that the three dimensional vector with coordinates $(S_1, S_2, S_3$), which we call the 3D Stokes vector, lies inside a sphere of radius $S_0$. This is the well known Poincar\'{e} sphere. Note that $S_0 = \Tr(\Phi)$ represents the total power of the beam.

The degree of polarization, $P^{(2)}$ of a two dimensional polarization matrix $\Phi$ is derived by writing $\Phi$ as the unique sum of two polarization matrices, one completely unpolarized (i.e. a multiple of the identity matrix), and one completely polarized (i.e. a rank 1 matrix) \cite{wolf07}. The degree of polarization is then the ratio of ``power" contained in the completely polarized matrix to the total power. It is given by
\begin{align}
P^{(2)}&=\sqrt{1-\frac{4\det(\Phi)}{\Tr[\Phi]^2}}.
\label{polarizationPhi}
\end{align}
Using \eq{phistokeseq} to write \eq{polarizationPhi} in terms of the Stokes parameters, we find that the degree of polarization is
\begin{align}
P^{(2)}&=\frac{\sqrt{S_1^2 + S_2^2 + S_3^2}}{S_0}.
\label{classicpolarization}
\end{align}
That is, the degree of polarization is simply the length of the 3D Stokes vector divided by the radius of the Poincar\'{e} sphere. Put differently, it is relative distance of our state from the origin of the Poincar\'{e} sphere. This makes intuitive sense and is a natural measurement, since the origin (where $P^{(2)}=0$) represents the completely unpolarized state, and the surface of the sphere (where $P^{(2)}=1$) represents the set of completely polarized states, and the value of $P^2$ for other states is ``linear" in the distance metric within the Poincar\'{e} sphere.
\subsection{Quantum Two Level system}

The most general quantum state is expressed in terms of the density matrix $\rho$, which contains all the statistically observable information of the state. This matrix is positive, Hermitian, and of unit trace. In the case of a quantum two level system, it is of dimension $2 \times 2$.
We can write $\rho$ as a linear combination of the identity and Pauli matrices as follows \cite{nielsenchuang}:
\begin{align}
\rho = \frac{1}{2}(I + r_1\sigma_1 + r_2\sigma_2 + r_3\sigma_3),
\label{rhoeq}
\end{align} 
where $\vec{r} = (r_1, r_2, r_3)$ is the well-known Bloch vector \cite{bloch}. Note that \eq{rhoeq} is in fact identical to \eq{phistokeseq}. Moreover, since $\rho$ is positive, we can also show that
\begin{align}
r_1^2 + r_2^2 + r_3^2 \le 1.
\label{blochsphere}
\end{align}
Therefore, the Bloch vector also lies within a (unit) sphere, known as the Bloch sphere. It is clear that in the two dimensional case, the quantum density matrix $\rho$ is analogous to the classical polarization matrix $\Phi$, and that the Bloch sphere is analogous to the Poincar\'{e} sphere.

The only mathematical difference between the two cases is one of scaling. To simplify the mathematics and make the link to the quantum case obvious, we set the density matrix $\rho$ to be the unit trace scaling of the polarization matrix $\Phi$. That is
\begin{align}
\rho &= \Phi/\Tr{[\Phi]}.
\end{align}
So $\rho$ is just the power-normalized version of $\Phi$. In the rest of this paper, we will only use $\rho$, keeping in mind that it applies for both quantum and classical cases. 

As a side note, we note that despite the identical mathematical formalism, the Bloch and Poincar\'{e} spheres differ in the underlying physical interpretation. If for example we take the quantum two level system to be the $\pm\frac{1}{2}$ spin states of a spin $\frac{1}{2}$ particle, then points on the Bloch sphere represent actual directions of spin in three dimensional space. In other words, each point on the Bloch sphere is an eigenstate of some (spin) angular momentum operator. 

The Poincar\'{e} sphere however does not have such a simple directional analogy. Its north pole represents right handed circularly polarized light, its south pole stands for left handed circularly polarized light, and its equatorial plane gives the linearly polarized states. Since the underlying classical beam of light is assumed be transverse, there is no longitudinal component. The relationship between the polarization states on the Poincar\'{e} sphere and three dimensional space is set by the direction of transverse propagation of the underlying light beam.

The photon being a spin 1 particle, can theoretically have a spin of 1, 0, or -1. However, the zero spin case represents longitudinal waves and is disallowed. Therefore we only have the $1$ and $-1$ spin states, that represent right and left circularly polarized light respectively. We can therefore conclude that contrary to the Bloch sphere, there are only two points on the Poincar\'{e} sphere that represent eigenstates of some spin angular momentum operator, the north and south poles, representing right and left circularly polarized light respectively.

\subsection{Polarization and Purity}

The origin of the Bloch sphere is the maximally mixed state whereas states on the surface of the sphere are pure states. Comparing this with the Poincar\'{e} sphere, where the origin is a completely unpolarized state and the surface contains completely polarized states, this suggests a direct analogue between quantum purity and and classical degree of polarization. 

However, the common measure of classical degree of polarization in \eq{classicpolarization} when expressed as a function of $\rho$ is given by the expression $\sqrt{1-4\det(\rho)}$, while the common measure of purity in quantum applications is given by $\Tr[\rho^2]$. Despite the clear physical analogy, there is a discrepancy in the measures used. This motivates us to analyze these and other measures of quantum purity and classical degrees of polarization that have been proposed in the literature. 

In the following sections, we will go through several such measures, and probe some of their properties and relationships, to find which one is appropriate in what situation.

\section{Measures of Purity for $N$ Dimensions}\label{s3:measures}
We proceed to introduce various measures of purity / polarization that have been suggested in the quantum mechanics and classical optics literature. Since purity is a property intrinsic to the density operator and invariant of the basis used, it should be invariant under unitary transformations. Therefore, one can always choose the basis where the density matrix is diagonal, and therefore, purity should be expressible as a function of the eigenvalues of $\rho$ alone, which we write as $\lambda_1 \ge \lambda_2 \ge ... \ge \lambda_N$, for an $N$ dimensional system.

We use the symbol $\Pi$ to denote the various measures of purity, with the appropriate subscript. If we write the purity as function of the eigenvalues, denoted by $\Pi(\lambda_1, ..., \lambda_N)$, then we require that it be a real-valued function that is scaled such that it takes values between $0$ and $1$. It should take value $1$ for a pure state and $0$ for the maximally mixed state; that is, $\Pi(1, 0, ..., 0) = 1$ and $\Pi(1/N, 1/N, ..., 1/N) = 0$, respectively. 

\subsection{Standard Purity}
In quantum information science, the common measure of purity  for a quantum state $\rho$ in an $N$ dimensional system is given by $ \Tr[\rho^2]$ \cite{nielsenchuang}. It takes a maximum value of $1$ for a pure state, and minimum value of $\frac{1}{N}$ for the maximally mixed state. This purity is sometimes scaled linearly so it varies between $0$ and $1$, giving the following expression, which we call standard purity:
\begin{align}
\Pi_s(\rho) \equiv \frac{N\Tr[\rho^2] -1}{N-1}.
\label{puritystd2}
\end{align}
In terms of eigenvalues, the standard purity is given by
\begin{equation}
\Pi_s(\lambda_1, ..., \lambda_N) = \frac{N\sum_{i=1}^N \lambda_i^2 - 1}{N-1}. \label{pisev}
\end{equation}

\subsection{Von Neumann Purity}
Shannon entropy is used in classical systems to quantify uncertainty about a random variable \cite{shannon}. Von Neumann entropy generalizes this to quantum systems, and is given by
\begin{equation}
S(\rho) \equiv -\Tr[\rho \log_2(\rho)]. 
\end{equation}
This measure quantifies the departure of a state from a pure state, i.e. its ``mixedness" \cite{nielsenchuang}. Note that the entropy of entanglement (a popular measure of entanglement for bipartite pure states) is defined to be the von Neumann entropy of one of the subsystems when the other subsystem is traced out, as discussed in section \ref{s5:entanglement}. This implies that the von Neumann entropy is a good measure of mixedness. Therefore, one can define another measure of purity, $\Pi_v(\rho) \in [0,1]$, based on the von Neumann entropy:
\begin{align}
\Pi_v(\rho) \equiv 1 + \frac{\Tr[\rho \log_2(\rho)]}{\log_2{N}}.
\label{purityvn}
\end{align}
Note that approximating the logarithm with its Taylor expansion, and ignoring higher order terms, we find a result that is a linear function of $\Pi_s$. That is, standard purity can be thought of as dervied from the Taylor approximation of von Neumann purity.

Expressed as a function of eigenvalues, von Neumann purity is given by
\begin{equation}
\Pi_v(\lambda_1, ..., \lambda_N) = 1 + \frac{1}{\log_2{N}}\sum_{i=1}^N \lambda_i \log_2(\lambda_i).
\label{pivev}
\end{equation}

If an eigenvalue $\lambda_k = 0$, we take $\lambda_k \log_2(\lambda_k) = 0$, since $\lim_{x\rightarrow 0^+} x \log(x) = 0$.
%
%
%

\subsection{Polarization Purity for N=2}

For the simple case of the two dimensional system, we have already seen that the classical degree of polarization is given by \eq{polarizationPhi}. Therefore the two dimensional polarization purity as a function of $\rho$ is
\begin{align}
P^{(2)}(\rho) &\equiv \sqrt{1-4\det(\rho)}.
\label{2dpuritypol}
\end{align} 
In terms of the two eigenvalues of $\rho$, this can be written as
\begin{align}
P^{(2)}(\lambda_1, \lambda_2) &= \sqrt{1 - 4 \lambda_1\lambda_2} \nonumber\\
&= \lambda_1-\lambda_2, \label{2dwolfpol}
\end{align} 
where in the last equality, we used the fact that $1 = (\lambda_1 + \lambda_2)^2$.
Finally, using \eq{classicpolarization} and expressing the Stokes parameters $S_i$ in terms of the Bloch vector elements $r_i = S_i/S_0$, i=1,2,3, we have
\begin{align}
P^{(2)}(\vec{r}) = \sqrt{r_1^2 + r_2^2 + r_3^2} = |\vec{r}|.
\label{2dradiuspol}
\end{align}
So we are left with three equivalent expressions for the polarization purity in two dimensions. Suppose we wish to generalize this measure of purity to $N\ge3$ dimensions. In principle there are an infinite number of ways to do this. However, only a handful of them have physical significance. In the following subsections, we discuss three possible generalizations, each one follows from one of the three equations (\ref{2dpuritypol}), (\ref{2dwolfpol}), and (\ref{2dradiuspol}). Although the three expressions for purity are identical in two dimensions, their respective extensions to higher dimensions differ, each forming its own measure.

In our generalizations, we pay  particular attention to the $N=3$ case, since it corresponds to classical polarization in three dimensions, a problem which has led to much debate in the literature \cite{friberg08, wolf04}. The idea of a three dimensional polarization is simple in principle. Rather than dealing with a beam propagating in one direction with polarization defined in the two dimensional transverse plane, one deals with an arbitrary electric field distribution in three dimensions. We may, for example, have classical light that contains longitudinal components and breaks the transversality condition. However, it is not clear what degree of polarization in this case means physically, leading to differing points of view.

\subsection{Barakat Heirarchy Measures of Purity}
In the case of an $N \times N$ density matrix $\rho$, Barakat has introduced a hierarchy of $N-1$ purity measures \cite{barakat83}. These measures are defined by first writing out the characteristic polynomial equation of $\rho$ as follows:
\begin{align}
\det{(\rho{-}\lambda I)} &= \lambda^N - C_1\lambda^{N-1} + C_2\lambda^{N-2} - ... + (-1)^NC_N \nonumber\\
&= 0.
\end{align}
The roots of this polynomial equation are by definition the eigenvalues of $\rho$. Each coefficient $C_k$ is the sum of all possible unique products of $k$ eigenvalues of $\rho$. That is
\begin{equation}
C_k = \sum_{1\le i_1 < ...< i_k \le N} \prod_{j=1}^k \lambda_{i_j}.
\label{ckdef}
\end{equation}
For example, if $N=3$, then
\begin{align}
C_1 &= \lambda_1 + \lambda_2 + \lambda_3 = 1, \nonumber\\
C_2 &= \lambda_1\lambda_2 +  \lambda_1\lambda_3 +  \lambda_2\lambda_3,\nonumber\\
C_3 &=  \lambda_1\lambda_2 \lambda_3 = \det{(\rho)}.
\end{align}
%
In fact $C_1=1$ and $C_N=\det(\rho)$ both hold for any $N$. Moreover, it can be shown that each $C_k$ is expressible in terms of $\Tr{(\rho^m)}$, for $m=2,...k$, or alternatively in terms of the first $k$ Casimir invariants of $\rho$ under the rotation group \cite{byrd}. For example, for any $N$, we have \cite{kimura}
\begin{align}
C_2 &= [1 - \Tr(\rho^2)]/2, \\
C_3 &= [1 - 3\Tr(\rho^2) + 2\Tr(\rho^3)]/6.
\label{ctotraces}
\end{align}
Therefore, the $C_k$ are invariant under change of coordinates. If $\rho$ is a pure state (i.e. has rank 1), then all the $C_k$ are zero, except for $C_1$ which is always unity. If $\rho$ is the maximally mixed state (all eigenvalues are $\frac{1}{N}$), then $C_k = \bigl(\begin{smallmatrix} N \\ k\end{smallmatrix} \bigr) \frac{1}{N^k}$ where $\bigl(\begin{smallmatrix} N \\ k\end{smallmatrix} \bigr)$ is the binomial coefficient. 

With this in mind and noting that $C_k$ coefficients themselves can be thought of as a measure of purity, Barakat then defines a hierarchy of measures of polarization given by $B_k^{(N)}(\rho)$ for $k = 2,... N$. Requiring that $B_2^{(2)}(\rho)$ collapse to $P^{(2)}(\rho)$ in \eq{2dpuritypol}, one defines
\begin{equation}
B_k^{(N)}(\rho) \equiv \sqrt{1-\small{\dbinom{N}{k}^{-1}N^kC_k}}, \hspace{10pt} k=2,...,N,
\label{bdef}
\end{equation}
The measure $B_k^{(N)}(\rho)$ takes the value zero for all $k$ in the maximally mixed (i.e. the fully unpolarized) state, and takes the value $1$ for all $k$ when $\rho$ is a pure (fully polarized) state. To get a feel for these measures, let us explore and simplify them using \eq{ctotraces} for some specific values of $N$ and $k$. For $N=2$, we have
\begin{equation}
B_2^{(2)}(\rho) = \sqrt{1 - 4\det(\rho)} = P^{(2)}(\rho),
\end{equation}
as we required. For $N=3$ we have
\begin{align}
B_2^{(3)}(\rho) &=  \sqrt{\big[3\Tr(\rho^2) - 1\big]/2}, \label{barakat23} \\
B_3^{(3)}(\rho) &= \sqrt{1 - 27\det(\rho)}\nonumber\\
&= \sqrt{1 - 27\lambda_1\lambda_2\lambda_3}.
\end{align}
For general $N$ we find
\begin{align}
B_2^{(N)}(\rho) &=  \sqrt{\frac{N \Tr(\rho^2) - 1}{N-1}} = \sqrt{\Pi_s(\rho)}, \label{barakat2n}\\
B_N^{(N)}(\rho) &= \sqrt{1 - N^N\det(\rho)}\nonumber\\
&= \sqrt{1 - N^N\lambda_1\lambda_2...\lambda_N}.\label{barakatnn}
\end{align}
Note the interesting relationship in \eq{barakat2n} where $B_2^{(N)}$ is simply the square root of the standard measure of purity $\Pi_s$. However, $B_N^{(N)}$ is unique among the measures we have so far, therefore we define \emph{Barakat's last measure} of purity as $\Pi_b$, given by
\begin{equation}
\Pi_b(\rho) \equiv B_N^{(N)}(\rho).
\end{equation}
We add $\Pi_b$ to our collection of measures which will be compared to other measures later in this paper. However, it must be mentioned that $\Pi_b$ has a serious shortcoming, in that $\Pi_b=1$ if any eigenvalue is zero. For example, it cannot distinguish between a pure state with eigenvalue spectrum $\{ 1, 0, ... , 0\}$, and a very mixed state with spectrum $\{\frac{1}{N}, \frac{1}{N}, ..., \frac{1}{N}, 0\}$. That is why Barakat measures are most effective when different levels of the hierarchy are used together.
%


\subsection{EDPW Purity}
Another measurement of purity is one proposed by Ellis, Dogariu, Ponomarenko, and Wolf \cite{wolf04, wolf04b}. It was presented as a measure of three dimensional polarization, however it has the same form for any dimension. The basic idea is measuring the total power in the fully polarized component. That is, one splits the $3 \times 3$ density matrix into a unique positive linear combination of the identity matrix, a rank 2 matrix with degenerate eigenvalues, and a rank 1 matrix. To illustrate, suppose $U$ is the unitary matrix that diagonalizes $\rho$, as per
\begin{equation}
U^\dag\rho U = \begin{bmatrix} \lambda_1 & 0 & 0 \\ 0 & \lambda_2 & 0 \\ 0 & 0 & \lambda_3 \end{bmatrix}.
\end{equation}
Then $\rho$ can be written as
\begin{align}
\rho = &(\lambda_1 {-} \lambda_2) \underbrace{U\begin{bmatrix} 1 & 0 & 0 \\ 0 & 0 & 0 \\ 0 & 0 & 0 \end{bmatrix}U^\dag}_{\substack{\text{fully polarized /} \\ \text{maximally pure} \\ \text{rank 1}}} + 
(\lambda_2 {-} \lambda_3) \underbrace{U\begin{bmatrix} 1 & 0 & 0 \\ 0 & 1 & 0 \\ 0 & 0 & 0 \end{bmatrix}U^\dag}_{\substack{\text{partially polarized} \\ \text{rank 2}}} \nonumber \\ 
&+ \lambda_3 \underbrace{U\begin{bmatrix} 1 & 0 & 0 \\ 0 & 1 & 0 \\ 0 & 0 & 1 \end{bmatrix}U^\dag}_{\substack{\text{unpolarized /} \\ \text{maximally mixed} \\ \text{rank 3}}}.
\label{rhodecompose}
\end{align}
Each of the coefficients $(\lambda_1 - \lambda_2)$,  $(\lambda_2 - \lambda_3)$, and $\lambda_3$ is positive, and the decomposition in \eq{rhodecompose} is unique. 
The \emph{EDPW purity}, denoted $\Pi_{edpw}$, is defined to be the ratio of the power in the fully polarized rank 1 matrix to the total power. That is, it is the ratio of the coefficient of the rank 1 matrix in the decomposition above to the sum of the eigenvalues, which is simply unity. Therefore, it is given by the simple expression
 \begin{equation}
\Pi_{edpw}(\lambda_1, ..., \lambda_N) = \lambda_1 - \lambda_2.
 \label{fullpoldef}
 \end{equation}
This is identical to $\eq{2dwolfpol}$ in two dimensions. In fact, this particular measure has the same form for any $N\ge2$, it is always the difference between the largest two eigenvalues. This can be seen by observing that \eq{rhodecompose} can be extended to any dimensionality without altering the first coefficient. 

The advantage of this measure is that it is physically meaningful. It is the fraction of the power that is completely polarized, and will be left unchanged if acted on by passive linear elements. That is, if we use some (hypothetical) three dimensional polarizers with the correct alignment, this fully polarized component is the only one that will remain unchanged.

However, when one considers the rank 2 component of the $\rho$ matrix, one sees that this component is not fully polarized, but neither is it fully unpolarized. This suggests that it must have some intermediate nonzero polarization of its own, and should make a contribution to the overall polarization / purity of the density matrix. Since $\Pi_{edpw}$ ignores the rank 2 component completely, it is not suitable as a measure of overall polarization, but is rather suited for measuring only the component of a field that is fully polarized. We clarify this in section \ref{subsecexample} with illustrative examples.  

\subsection{SSKF Purity} 
The measure of purity due to Set\"{a}l\"{a}, Shevchenko, Kaivola, and Friberg \cite{friberg02} starts by writing the $3 \times 3$ density matrix $\rho$ as a linear combination of some basis matrices in an expression similar to \eq{rhoeq}. We write $\rho$ as
\begin{equation}
\rho = \frac{1}{3}I + \frac{1}{\sqrt{3}}\sum_{i=1}^8r_i\hat{G}_i,
\label{3drhoeq}
\end{equation}
where $I$ is the identity matrix, and $\hat{G}_i, i=1-8$ are the popular three dimensional analogue of the Pauli matrices, the Gell-Mann matrices \cite{gellmann}, shown in section \ref{gellmannappendix} of the appendix. We have modified the coefficients from those in ref. \cite{friberg02} to ease calculation. The eight coefficients $r_i$ together form a (generalized) Bloch vector $\vec{r}$.  One can use \eq{3drhoeq} together with the orthogonality and tracelessness of Gell-Mann matrices to show that
\begin{equation}
\Tr{[\rho^2]} = \frac{1}{3} + \frac{2}{3}|\vec{r}|^2.
\label{sqrtraceproperty}
\end{equation}
The density matrix property $\Tr{[\rho^2]} \le 1$ together with \eq{sqrtraceproperty} imply that $\sum_{i=1}^8r_i^2 = |\vec{r}|^2 \le 1$. That is, the Bloch vector $\vec{r}$ lies inside an eight dimensional hypersphere of unit radius. We then define the \emph{SSKF purity}, denoted $\Pi_{sskf}$, in a manner analogous to \eq{classicpolarization} and \eq{2dradiuspol} as the length of the Bloch vector, i.e. the radial distance from the origin in this hypersphere. It is given by
 \begin{equation}
 \Pi_{sskf}(\vec{r}) \equiv \Big[ \sum_{i=1}^8r_i^2 \Big]^{\frac{1}{2}} = |\vec{r}|.
 \label{pirgellmann}
 \end{equation}
We may alternatively call this measure the radial purity, emphasizing that it gives the length of a radial Bloch vector in a hypersphere. Equivalently, one can also solve \eq{sqrtraceproperty} for $|\vec{r}|$ to write the SSKF purity as a direct function of $\rho$: 
 \begin{equation}
 \Pi_{sskf}(\rho) = \sqrt{\big[3\Tr(\rho^2) - 1\big]/2},
\label{sskfrho}
 \end{equation}
 or as a function of the eigenvalues:
 \begin{equation}
 \Pi_{sskf}(\lambda_1, \lambda_2, \lambda_3) = \sqrt{\frac{1}{2}\Big[3(\lambda_1^2 + \lambda_2^2 + \lambda_3^2) - 1 \Big]}.
 \label{sskfevalues}
 \end{equation}
Note that \eq{sskfrho} above is identical to \eq{barakat23}, and therefore  $\Pi_{sskf}(\rho) \equiv B_2^{(3)}(\rho)$ for $N=3$. 
To generalize this to general dimensionality $N$, we write 
\begin{equation}
\rho = \frac{1}{N}I + \frac{1}{\sqrt{N}}\sum_{i=1}^{N^2-1}r_i\hat{Q}_i,
\label{Ndrhoeq}
\end{equation}
where $\hat{Q}_i$ are traceless operators that form a basis for $SU(N)$, and satisfy the orthogonality relation $\Tr{[\hat{Q}_i\hat{Q}_j]}=(N-1)\delta_{ij}$. The Bloch vector $\vec{r}$ has $N^2-1$ entries $r_i$. Squaring \eq{Ndrhoeq} and taking the trace we find
\begin{equation}
\Tr{[\rho^2]} = \frac{1}{N} + \frac{N-1}{N}|\vec{r}|^2.
\label{sqrtracepropertyN}
\end{equation}
From this, we can conclude that for any dimensionality $N$
\begin{align}
\Pi_{sskf}(\rho) &\equiv |\vec{r}| \\
&= \sqrt{\frac{N\Tr[\rho^2] -1}{N-1}}.\label{pirgeneraln} \\
&=B_2^{(N)}(\rho) \\
&= \sqrt{\Pi_s(\rho)}.
\end{align} 

So we find that the SSKF / radial purity is equivalent to Barakat's second measure and the square root of the standard measure. This is a very interesting result since each of these measures was ostensibly derived in a different manner. It shows that we really have fewer measures than may initially seem.

Reverting back to the $N=3$ case, the picture that has formed seems a natural generalization of the familiar Poincar\'{e}/Bloch sphere. The centre of the eight dimensional hypersphere still represents the totally unpolarized (or maximally mixed) state, and the states on the surface of the hypersphere are totally polarized (pure). 

There is one essential difference however, that undermines the elegance and simplicity of this picture. In the case $N=2$, all states in or on the Bloch sphere represent valid physical states and a positive density matrix. However, in dimensionality $N=3$ or higher, it has been be shown that the physical constraint of positivity on the density matrix restricts the set of valid states to an irregular convex region that is a proper subset of the enclosing hypersphere. This physical region touches the surface of the enclosing hypersphere only in some places (where the fully polarized states lie) \cite{byrd, kimura}. That is, many states within the hypersphere and on its surface are unphysical since they would create density matrices that are not positive. 

\begin{figure}[h!]
 \centerline{\includegraphics[width=\columnwidth]{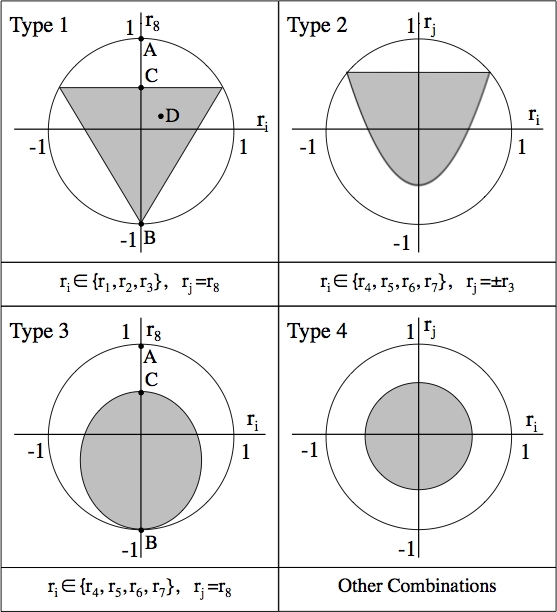}} 
\caption{Classes of cross section of the eight dimensional space in which the generalized Bloch vectors live, based on a figure by Kimura \cite{kimura}. In each diagram, the shaded region represents the allowable states, while the outer circle is a cross section of the enclosing hypersphere. The pure states are where the shaded region touches the outer circle. Points A, B, C, D are specific states we examine.}
\label{kimuradiag}
\end{figure}

Figure \ref{kimuradiag} shows all possible two dimensional cross sections of the eight dimensional hypersphere. The shaded regions represent physical polarization/density matrices. Note that in fact most of the volume inside the hypersphere will be composed of disallowed unphysical states. 

 %
%
A state lying on the surface of the hypersphere is a necessary but insufficient condition for it to represent a pure state, for it is unphysical if it is not on the border of the allowable region. States anywhere on the border of allowable region must have at least one zero eigenvalue. A generic allowable state does not necessarily lie on a straight line between the maximally mixed state and a pure state as in the case of the three dimensional Bloch sphere. This must have been the case, since a positive $3 \times 3$ matrix in general cannot just be written as a linear combination of the identity matrix (maximally mixed) and a rank $1$ matrix (pure), there is generally a rank $2$ component as shown in \eq{rhodecompose}. 

To illustrate these features, let us examine the states represented by points $A, B, C$ and $D$ in fig. (\ref{kimuradiag}). These four points are given by the following Bloch vectors:
\begin{align}
\vec{r}_A &= (0, 0, 0, 0, 0, 0, 0, 1), \nonumber\\
\vec{r}_B &= (0, 0, 0, 0, 0, 0, 0, -1), \nonumber\\
\vec{r}_C &= (0, 0, 0, 0, 0, 0, 0, 1/2), \nonumber\\
\vec{r}_D &= (0, 0, \sqrt{3}/8, 0, 0, 0, 0, 1/8).
\end{align}
Using \eq{3drhoeq}, we can construct the corresponding density matrices, and we find
\begin{align}
\rho_A &= \begin{bmatrix} 2/3 & 0 & 0 \\ 0 & 2/3 & 0 \\ 0 & 0 & -1/3 \end{bmatrix}, \hspace{10pt} &\rho_B &= \begin{bmatrix} \ \ 0\ \ & \ 0 \ & \ \ 0 \ \ \\ 0 & 0 & 0 \\ 0 & 0 & 1 \end{bmatrix}, \nonumber\\
\rho_C &= \begin{bmatrix} 1/2 & 0 & 0 \\ 0 & 1/2 & 0 \\ 0 & 0 & \ \ 0 \ \ \end{bmatrix},  \hspace{10pt}&\rho_D &= \begin{bmatrix} 1/2 & 0 & 0 \\ 0 & 1/4 & 0 \\ 0 & 0 & 1/4 \end{bmatrix}.
\label{examplematrices}
\end{align}
We see that $\rho_A$ is not positive, and therefore unphysical, despite lying on the unit hypersphere, since it is not in or on the border of the allowable region. This illustrates the breakdown of the analogy with the two dimensional Bloch sphere addressed above. The matrix $\rho_B$ however is positive and therefore physical. Given that it is physical, we can see that it must be a pure state since it lies on the surface of hypersphere, and indeed it is. The matrix $\rho_C$ is also physical, and has a single zero eigenvalue, which is expected since it is at the boundary of allowable states. If it were to move slightly outside the boundary, the zero eigenvalue would become negative and therefore unphysical. The state given by $\rho_D$ is a typical state inside the allowable region.

One may suggest that these properties are a result of an artificial asymmetry of the Gell-Mann matrices (in particular $G_3$ and $G_8$), and may be avoided if we opt for a different basis set of matrices for $SU(3)$. However, this is not true, and the qualitative properties illustrated above are intrinsic to the $N=3$ case, and still hold even if one exchanges the Gell-Mann matrices for a different basis set with the same basic properties of Hermiticity, tracelessnesss, and orthogonality. 

To see this, note the following basis-independent property: the surface of the hypersphere is seven dimensional, and pure states only form a three dimensional surface. This implies that, independent of the choice of the basis, the pure states form a very small part of the surface of the generalized Bloch hypersphere. Most states on the hypersphere surface will be analogous to $\rho_A$ above, that is they will be unphysical due to violation of positivity. 

For general $N$, the enclosing hypersphere is of dimension $N^2-1$, and its surface of dimension $N^2-2$. The space of pure states is only of dimension $N$. Only in the case $N=2$ do we have the dimensions of the surface and of the pure state space coinciding, giving us the simple properties of the conventional Bloch sphere.

Yet despite the loss of the simple geometry of a filled hypersphere, the SSKF / radial purity still, in some sense, quantifies the distance of the state from the maximally mixed state. Furthermore, if we suppose that the system of interest involves depolarizing channels, a popular type of quantum noise channel \cite{king}, we find that $\Pi_{sskf}$ satisfies an intuitive depolarization criterion, making it the most convenient and logical purity measure. We discuss this in more detail in appendix \ref{appendix:sskfdepol}.

\section{Comparing Purity Measures for Three Dimensions}\label{s4:comparison}

\subsection{Graphical Comparison}\label{s4:graphcomparison}

Thus far, we have discussed five contending measures of purity for $N=3$ dimensions: the standard purity $\Pi_s$, the von Neumann purity $\Pi_v$, Barakat's last measure $\Pi_b$, the EDPW purity $\Pi_{edpw}$, and the SSKF purity measure $\Pi_{sskf}$. Since the standard purity $\Pi_s$ is just the square of the SSKF purity $\Pi_{sskf}$, we ignore the former and only include the latter in our comparison. 

\begin{figure}[h!]
 \centerline{\includegraphics[width=203pt]{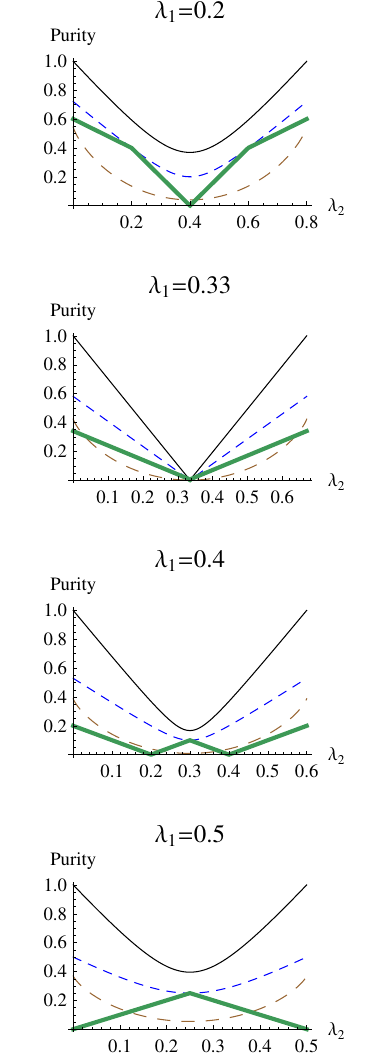}} 
\caption{Values of purity measures for various eigenvalues of a three dimensional density matrix. Each graph has fixed $\lambda_1$, with $\lambda_2$ against the horizontal axis, and $\lambda_3 \equiv 1 - \lambda_1 - \lambda_2$. In each graph, the upper solid curve in black is Barakat's last measure $\Pi_b$, the upper dashed blue curve is the SSKF purity $\Pi_{sskf}$, the lower dashed brown curve is the von Neumann purity $\Pi_v$, and the lower solid green curve is the EDPW purity $\Pi_{edpw}$.}
\label{puritygraphs}
\end{figure}

To compare the four remaining measures, we set $N=3$, and recall that the purity will only be a function of the eigenvalues $\lambda_1, \lambda_2,$ and $\lambda_3 = 1-\lambda_1-\lambda_2$, leaving us with only two degrees of freedom. In figure \ref{puritygraphs}, we plot the various measures of purity against $\lambda_2$, for some fixed values of $\lambda_1$. Note the interesting point in the second graph where all the purity measures are zero, this is the maximally mixed state ($\lambda_1 = \lambda_2 =\lambda_3 = \frac{1}{3}$).
%
%
%

Examining figure \ref{puritygraphs}, we first note that in the graphs above, $\Pi_b$, $\Pi_{sskf}$ and $\Pi_v$ behave similarly. That is, comparing the purity of any two states within the same graph, these three measures agree which state is more pure. The values of these measures all increase together, decrease together, and have extrema at the same eigenvalues. Derivatives of their three curves always have the same sign. In appendix \ref{monoticeq}, we show that this behaviour of these three measures will be seen for any dimensionality $N$, provided we only vary two eigenvalues and not more. 


We also observe that $\Pi_{edpw}$ often, but not always, yields results opposite to those of all the other measures. That is, it sometimes disagrees with other measures in deciding which of two states is purer. This suggests that it is measuring something entirely different, and can be better understood through examples.

\subsection{Illustrative Examples}\label{subsecexample}
Recall state C with eigenvalue spectrum $\{1/2, 1/2, 0\}$ and state D with spectrum $\{1/2, 1/4, 1/4\}$, with their respective density matrices $\rho_C$ and $\rho_D$ defined in \eq{examplematrices}. 
%

Suppose we wish to use one of our measures of purity to find which density matrix $\rho_C$ or $\rho_D$, is more pure. If for example we use the measure of purity $\Pi_{sskf}$, we find that state $\rho_C$ has higher purity than state D. If we use $\Pi_{edpw}$ we find the opposite, state D is higher in purity. To get a more comprehensive idea, we also introduce state E with eigenvalue spectrum $\{3/4, 1/8, 1/8 \}$ and state F with spectrum $\{2/3, 1/6, 1/6 \}$. In table \ref{puritytablevalues}, we evaluate the measures $\Pi_{sskf}$, $\Pi_{edpw}$, $\Pi_b$ and $\Pi_v$ for all of these states.
\begin{table}
{\small \begin{tabular}{c || c | c | c | c | c | c }  
  & P &  E &  F &  C &  D & M  \\ \hline
  &  $(1, 0, 0)$ & $(\frac{3}{4}, \frac{1}{8}, \frac{1}{8})$ & $(\frac{2}{3}, \frac{1}{6}, \frac{1}{6})$ & $(\frac{1}{2}, \frac{1}{2}, 0)$ & $(\frac{1}{2}, \frac{1}{4}, \frac{1}{4})$ & $(\frac{1}{3}, \frac{1}{3}, \frac{1}{3})$  \\
  \hline
$\Pi_{sskf}$	  & 1 & 0.625 & 0.5     & 0.5     & 0.25   & 0\\
$\Pi_{edpw}$ & 1 & 0.625 & 0.5     & 0        & 0.25    & 0\\
$\Pi_{b}$ 	  & 1 & 0.827 & 0.707 & 1        & 0.395 & 0\\
$\Pi_{v}$ 	  & 1 &  0.330& 0.210 & 0.369 & 0.054 & 0\\
\end{tabular}
}
\caption{\label{puritytablevalues} The purity of the states C, D, E, and F introduced above, as well as the pure (P) and maximally mixed (M) states. Purity is evaluated by four different measures: $\Pi_{sskf}$, $\Pi_{edpw}$,  $\Pi_b$ and $\Pi_v$. The eigenvalue spectrum $(\lambda_1, \lambda_2, \lambda_3)$ of each state in the second row. The columns are ordered from purest to most mixed according to the $\Pi_{sskf}$ measure.} 
\end{table}
The most striking feature of table \ref{puritytablevalues} is that no two of the measures agree on the ordering of the states from purest to most mixed. To help us reason more clearly, we note that in general, the more mixed a state is, the closer all the eigenvalues are to each other, with the extreme case being the maximally mixed state where all eigenvalues are equal (to $1/N$). The purer a state is the more a small number of eigenvalues should stand out, with pure states having a single eigenvalue equal to unity and as far as possible from the rest, which are all zero.

Restricting ourselves to the states C and D, we can now reason that each of the density matrices $\rho_C$ and $\rho_D$ have two identical eigenvalues (a ``mixed" property), but in $\rho_C$, the third distinct eigenvalue is further away from the identical two than in $\rho_D$ (that is, $|0 - 1/2| > |1/2 - 1/4|$), therefore $\rho_C$ should be more pure. We can also reason that since both states have one eigenvalue of 1/2, then they are equal in this respect, and the other two eigenvalues should be the deciding factor in which state is more pure. The remaining eigenvalues for matrix $\rho_D$ are 1/4, 1/4, these are identical (more mixed), and for matrix $\rho_C$ are 1/2, 0, these are as different as possible. So we expect that $\rho_C$ must have higher overall purity. Therefore it seems $\Pi_{sskf}$ is more suited for the general idea of overall purity than $\Pi_{edpw}$.  

However, suppose instead of overall purity, we are interested in the component of the density matrix that is fully polarized (i.e. the component that can be acted upon by a hypothetical three dimensional polarizer and remain unchanged). We see that we can write
\begin{equation}
\rho_D = \frac{1}{4}\begin{bmatrix} 1 & 0 & 0 \\ 0 & 0 & 0 \\ 0 & 0 & 0 \end{bmatrix} + \frac{1}{4}\ I.
\end{equation}
That is, $\rho_D$ has a nontrivial fully polarized component, the magnitude of which will be given by  $\Pi_{edpw}(\rho_D) = 1/4$. The matrix $\rho_C$ however cannot be decomposed in this way, has no fully polarized component, and therefore $\Pi_{edpw}(\rho_C) = 0$. We conclude that the choice of a more suitable measure of purity depends on what one is interested in measuring, though $\Pi_{sskf}$ seems more suitable for general purposes. Since the standard purity $\Pi_{s}$ is simply the square of the latter, it can be used as quick and simple way to measure purity, and its ubiquity in quantum information seems justified.

\subsection{Relationship between SSKF Purity $\Pi_{sskf}$ and EDPW Purity $\Pi_{edpw}$}
We have already discussed the properties, strengths and weaknesses of $\Pi_{sskf}$ and $\Pi_{edpw}$. It is of interest to find a simple relationship between the two measures with aid of a pair of suitably defined variables. The following analysis is similar to results by Sheppard \cite{sheppard}. In the $N=3$ case there are only two degrees of freedom in setting the eigenvalues (since they must sum to unity). We define the variables $x$ and $y$ as follows:
\begin{align}
x &\equiv \Pi_{edpw} = \lambda_1 - \lambda_2, \hspace{10pt} \nonumber \\
y &\equiv 3(\lambda_1 + \lambda_2 - 2/3) = 1 - 3\lambda_3.
\end{align}
Physically, $x$ is the EDPW purity, i.e. the fraction of the power that is in the fully polarized component. and $y$ can be thought of as the fraction of power that is \emph{not} in the completely unpolarized / mixed component. In other words, $x$ represents the power in the rank 1 component of the density matrix, while $y$ represents power in the rank 1 or rank 2 components, i.e. the power not in the rank 3 component. Both $x$ and $y$ vary between $0$ and $1$, with the condition that $y \ge x$. This latter inequality can be seen from
\begin{align}
y - x &= 1- \lambda_1 + \lambda_2 - 3\lambda_3 \nonumber\\
&=2\lambda_2 - 2\lambda_3 \ge 0,
\end{align}
where in the second equality we used  $1=\lambda_1 + \lambda_2 + \lambda_3$, and the last inequality we noted that  $\lambda_2\ge\lambda_3$. Then we can express the eigenvalues in terms of $x$ and $y$:
\begin{align}
\lambda_1 &= \frac{1}{3} + \frac{1}{2}(\frac{y}{3} + x), \\
\lambda_2 &= \frac{1}{3} + \frac{1}{2}(\frac{y}{3} - x), \\
\lambda_3 &= \frac{1}{3} - \frac{y}{3}.
\end{align}
We can make use of these expressions to show that $\lambda_1^2 + \lambda_2^2 + \lambda_3^2 = (2 + 3x^2 + y^2)/6$. Using this result along with \eq{sskfevalues}, $\Pi_{sskf}$ is expressible as
\begin{align}
\Pi_{sskf} = \frac{1}{2}\sqrt{3x^2 + y^2}.
\label{sskf:xy}
\end{align}
This tells us that $\Pi_{sskf}$ includes the purity from $x$ (i.e. $\Pi_{edpw}$) plus an additional component from $y$. Note that for a given $x$, the minimum value $y$ can take is $x$, in which case we see from \eq{sskfevalues} that $\Pi_{sskf} = x = \Pi_{edpw}$. This is expected, since for $y$ to equal $x$, this means there is no power in the rank 2 component, and all the polarized power is in the rank 1 component, so both measures agree. 

\section{Relation to Entanglement Measures}\label{s5:entanglement}

Erwin Schr\"{o}dinger first pointed out the uniquely quantum phenomenon of entanglement in his seminal 1935 paper with Max Born \cite{schrodinger}. At that time, entanglement was poorly understood, and subject to paradoxes, such as the famous Einstein-Podolsky-Rosen thought experiment \cite{epr}, which was subsequently analyzed by John Bell, leading him to his famous inequalities \cite{bell}. Though much of the same fundamental mystery of quantum entanglement remains today, we have at least a large array of potential tools to measure it \cite{horodecki}. In this section, we propose yet another potential approach with which to measure entanglement, namely measuring the purity of a subsystem through our various purity measures.

The entanglement of a bipartite system is directly related to the purity of a subsystem once the other subsystem has been traced out. For example, say we have a bipartite system of two qubits, $A$ and $B$, given by the Bell state
\begin{equation}
\ket{\Phi}_{AB} = \frac{\ket{00} + \ket{11}}{\sqrt{2}}.
\end{equation}
This system is maximally entangled. If we define $\rho_A$ as the improper density matrix of the first qubit once the second one has been traced, we get
\begin{align}
\rho_A &\equiv \Tr_B{[\ket{\Phi}\bra{\Phi}]} \nonumber\\
&=\begin{pmatrix} 1/2 & 0 \\ 0 & 1/2\end{pmatrix}.
\end{align}
 
Note that $\rho_A$ is maximally mixed. If we had traced out system $A$ and kept $\rho_B$ it would have been identical. Moreover, if $\ket{\Psi}_{AB}$ where a separable state, then $\rho_A$ would have been a pure rank 1 matrix. So, we see that maximal entanglement leads to maximal mixedness in the subsystem, and no entanglement leads to a pure subsystem state. This argument suggests that the mixedness (one minus the purity) of a subsystem is a good measure of entanglement of the whole system. 

A common measure of entanglement for a bipartite system is entropy of entanglement $E$ \cite{greenberger}. It is defined as the von Neumann entropy $S$ (i.e. a measure of mixedness) of a subsystem once the other has been partially traced out. That is, it can be written as
\begin{equation}
E(\ket{\Psi}) = 1-\Pi_v[\Tr_B(\ket{\Psi}\bra{\Psi})].
\label{entanglepurity}
\end{equation}
What if we replace $\Pi_v$ in \eq{entanglepurity} with another measure of purity, say $\Pi_{sskf}$ or $\Pi_b$? This would give rise to another class of entanglement measures with different properties, which could possibly be more relevant for some applications.

For example we consider a bipartite system with $N=3$, i.e. a system of two qutrits. Such a system has been studied by some authors, and even geometric descriptions developed to help visualize its entanglement \cite{qutrit}. Say we had the following two states:
\begin{align}
\ket{\Psi_C} &= \frac{\ket{00} + \ket{11}}{\sqrt{2}},\\
\ket{\Psi_D} &= \frac{1}{\sqrt{2}}\ket{00} + \frac{1}{2}\ket{11} + \frac{1}{2}\ket{22} \nonumber\\
&=\frac{\sqrt{2}-1}{2}\ket{00} + \frac{\sqrt{3}}{2} \Big( \frac{\ket{00} + \ket{11} + \ket{22}}{\sqrt{3}}\Big).
\end{align}
If we partially trace out one subsystem in each of these two states, we are left with the density matrices $\rho_C$ and $\rho_D$ given by \eq{examplematrices} in the previous section. Then we can revisit the discussion in section \ref{subsecexample} regarding which measures of purity are more suitable, $\Pi_{sskf}$ or $\Pi_{edpw}$. The question of which is more pure, $\rho_C$ or $\rho_D$, can be asked differently: which is less entangled, $\ket{\Psi_C}$ or $\ket{\Psi_D}$? 

In this case, we can further appreciate why $\Pi_{edpw}$ favoured $\rho_D$ as more pure, because it favours $\ket{\Psi_C}$ as more entangled. Note that $\ket{\Psi_C}$ is a two dimensional Bell state in a three dimensional system, and one can think of it in a very specific sense as more entangled than $\ket{\Psi_D}$ since it is equal to a Bell state (even if it is of a lower dimension). One may also reason that $\ket{\Psi_C}$ should have higher entanglement since it has no separable component, whereas $\ket{\Psi_D}$ can be written as a combination of the separable state $\ket{00}$, and the maximally entangled qutrit state as shown above. Note that the latter is a linear combination of non-orthogonal states, and therefore the squares of the coefficients $\frac{\sqrt{2}-1}{2}$ and $ \frac{\sqrt{3}}{2}$ do not add to unity.

However, we can conclude $\ket{\Psi_D}$ is more entangled than $\ket{\Psi_C}$ (and therefore $\rho_C$ purer than $\rho_D$), since both have the same coefficient for the $\ket{00}$ state, yet $\ket{\Psi_D}$ has both a $\ket{11}$ and $\ket{22}$ component while  $\ket{\Psi_C}$ only has a $\ket{11}$ with no entanglement in the third level whatsoever. This definition of entanglement is the one of interest for practical purposes.

All of this of course assumes the bipartite state $\ket{\Psi}_{AB}$, is pure. In general, this state can itself be mixed and is written as a density matrix $\rho_{AB}$, which complicates the situation, and gives rise to the large and growing number of entanglement measures in the literature \cite{horodecki}. Furthermore, it is not immediately clear how this idea of using purity to measure entanglement may be generalized to measures of tripartite entanglement between three systems, or multipartite entanglement where an arbitrary number of systems is involved. Most likely, it would involve a type of average of the purities of each each subsystem individually, obtained by tracing out all the other systems. This may open the door to new interesting measures of entanglement applicable in different situations.

Entanglement can also be related to purity / polarization in other subtle ways. For example, the Schmidt decomposition can be used to decompose entangled states to a unique positive sum of separable states \cite{schmidt, ekertknight}. This has been exploited by some authors to create a measure of polarization based on the Schmidt decomposition \cite{qianeberly}.


\section{Conclusion}
We showed that quantifying quantum purity for an $N$ level system is equivalent mathematically to quantifying the degree of classical polarization in $N$ dimensions. Then we described and analyzed different measures of purity, finding interesting properties and strength and weaknesses of each.

In the more common case of measuring overall purity, the SSKF / radial purity $\Pi_{sskf}$ seems the strongest option, since it is most consistent with depolarizing channels, commonly used quantum noise channels. The measure is also motivated through a simple geometric analogy of a generalized Bloch sphere, which still provides insights despite some limitations in higher dimensions. The standard purity $\Pi_s$ in particular is the simplest to use and the most common in the quantum information literature. It turns out to be just the square of $\Pi_{sskf}$, and therefore inherits its validity as a measure of purity.

Barakat's last measure of purity $\Pi_b$ is easy to compute, but becomes useless as soon as one of the eigenvalues approaches zero. The von Neumann purity $\Pi_v$ is also interesting due to its connection to entropy, but it has few useful properties. 

If instead we are interested only in the component that is fully polarized, then the EDPW purity $\Pi_{edpw}$ is a more suitable measure. It will yield the strength of only the fully polarized part, discarding other components. It can also be shown that for $N=3$, $\Pi_{sskf}$ and $\Pi_{edpw}$ are related in a simple manner once we add a variable to represent the second degree of freedom.

Moreover, there is a direct relationship between the entanglement of a pure bipartite state and the purity of one of its subsystems once the other subsystem has been traced out. This can be used to give insight into measures of entanglement, and possibly create new entanglement measures based on various measures of purity. 


\section*{Acknowledgements}
This work was funded by the Natural Sciences and Engineering Research Council of Canada (NSERC), and the Ontario Ministry of Training, Colleges, and Universities.

\newpage
\appendix

\section{Gell-Mann Matrices}\label{gellmannappendix}

The Gell-Mann matrices $(G_i, i=1,...,8)$ are the most widely used set of generators for the group of special unitary $3\times 3$ matrices, $SU(3)$ \cite{gellmann}. They are given by
\begin{align}
G_1 &= \begin{bmatrix} 0 & 1 & 0 \\ 1 & 0 & 0 \\ 0 & 0 & 0 \end{bmatrix},  \hspace{10pt}&G_2 &= \begin{bmatrix} 0 & -i & 0 \\ i & 0 & 0 \\ 0 & 0 & 0 \end{bmatrix}, \nonumber\\
G_3 &= \begin{bmatrix} 1 & 0 & 0 \\ 0 & -1 & 0 \\ 0 & 0 & 0 \end{bmatrix},  \hspace{10pt}&G_4 &= \begin{bmatrix} 0 & 0 & 1 \\ 0 & 0 & 0 \\ 1 & 0 & 0 \end{bmatrix}, \nonumber\\
G_5 &= \begin{bmatrix} 0 & 0 & -i \\ 0 & 0 & 0 \\ i & 0 & 0 \end{bmatrix},  \hspace{10pt}&G_6 &= \begin{bmatrix} 0 & 0 & 0 \\ 0 & 0 & 1 \\ 0 & 1 & 0 \end{bmatrix}, \nonumber\\
G_7 &= \begin{bmatrix} 0 & 0 & 0 \\ 0 & 0 & -i \\ 0 & i & 0 \end{bmatrix},  \hspace{10pt}&G_8 &= \frac{1}{\sqrt{3}}\begin{bmatrix} 1 & 0 & 0 \\ 0 & 1 & 0 \\ 0 & 0 & -2 \end{bmatrix}. 
\end{align}
They are all Hermitian, traceless, and satisfy the orthogonality relation $\Tr{[G_iG_j]} = 2\delta_{ij}$, where $\delta_{ij}$ is the Kronecker delta. However they are not unitary like the Pauli matrices.

\section{Depolarizing Channels as a Criteria}\label{appendix:sskfdepol}

Consider a depolarizing channel, an important type of quantum noise \cite{nielsenchuang}. It is a transformation which \emph{depolarizes} the input quantum state (i.e. replaces it with $I/N$) with probability $1-p$, and leaves it unchanged with probability $p$. The action of this channel on the density matrix is given by the following superoperator \cite{king}:
\begin{align}
\mathcal{E}(\rho) = (1-p)\frac{I}{N} + p\rho.
\label{depoldef} 
\end{align}
Squaring \eq{depoldef} and taking the trace, we have
\begin{align}
\Tr[\mathcal{E}(\rho)^2] &= \frac{(1-p)^2}{N} + \frac{2p(1-p)}{N} + p^2\Tr[\rho^2] \nonumber\\
&= \frac{1-p^2}{N} + p^2\Tr[\rho^2].
\label{tracesqrdepol}
\end{align}
Applying the SSKF polarization measure of purity in \eq{pirgeneraln} to $\mathcal{E}(\rho)$, we have
%
\begin{align}
\Pi_{sskf}^2(\mathcal{E}(\rho)) &= \frac{N\Tr[\mathcal{E}(\rho)^2] - 1}{N-1}\nonumber\\
&= \frac{Np^2\Tr[\rho^2] +1 -p^2 - 1}{N-1}\nonumber\\
&= p^2\frac{N\Tr[\rho^2] - 1}{N-1}\nonumber\\
&= p^2\Pi_{sskf}^2(\rho).
\end{align}
where in the second line we made use of \eq{tracesqrdepol}. Taking the square root of both sides, we have the simple result
\begin{equation}
\Pi_{sskf}(\mathcal{E}(\rho))=p\Pi_{sskf}(\rho). \label{pirdepol}
\end{equation}
We observe that \eq{pirdepol} has a very simple and intuitive form, showing that the purity simply scales down by a factor of $p$ after the state passes through the depolarizing channel. This intuitive relationship does not hold for other measures of purity, even ones whose partial derivatives have the same sign as $\Pi_{sskf}$, shown in the next section \ref{monoticeq}. This suggests $\Pi_{sskf}$ is a measure with more physical meaning, and more relevant whenever depolarizing channels are in effect, as they often are.

\section{Partial Derivatives and Agreement of Purity Measures}\label{monoticeq}

The graphical comparison in section \ref{s4:graphcomparison} shows that the measures $\Pi_b$, $\Pi_{sskf}$ and $\Pi_v$ behave similarly; their derivative always has the same sign, and one is tempted to conclude that they will always agree which of any two states is more pure. It is of interest to ask if this will aways be true. To gain some insight into this question, we first examine the signs of the partial derivatives of the various measures.

Assume we have an $N$ dimensional state, with eigenvalues $\lambda_1, ..., \lambda_{N}$, with the first $N-1$ eigenvalues independent, and $\lambda_N$ a dependent variable satisfying
\begin{equation}
\lambda_N = 1- \sum_{j=1}^{N-1} \lambda_j.
 \end{equation}
Then by \eq{pisev}, \eq{pivev}, and \eq{barakatnn}, we have
\begin{align}
\Pi_s &= \frac{N[\sum\lambda_j^2 + (1-\sum\lambda_j)^2] - 1}{N-1}, \\
\Pi_v &= 1 + \frac{\sum\lambda_j\log_2\lambda_j + (1-\sum\lambda_j)\log_2(1-\sum\lambda_j)}{\log_2N}, \\
\Pi_b^2 &= 1 - N^N\lambda_1\lambda_2...\lambda_{N-1}(1-\sum\lambda_j),
\end{align}
where all sums over $j$ in this section run from $1$ to $N-1$. We have squared $\Pi_b$ since it does not affect the sign of the derivative, and makes the calculation more tractable. Also, since $\Pi_s = \Pi_{sskf}^2$, the properties we find for  $\Pi_s$ will also apply to $\Pi_{sskf}$.

Keeping in mind that $\frac{\partial\lambda_N}{\partial\lambda_i}=-1$ for $i=1, ..., N$, we can compute the partial derivatives $\frac{\partial\Pi_s}{\partial\lambda_i}$, $\frac{\partial\Pi_v}{\partial\lambda_i}$, and $\frac{\partial\Pi_b^2}{\partial\lambda_i}$ as follows:
%
%
\begin{align}
\frac{\partial\Pi_s}{\partial\lambda_i} &=  \frac{N[2\lambda_i - 2(1-\sum\lambda_j)]}{N-1} \nonumber\\ 
&=\frac{2N[\lambda_i - \lambda_N]}{N-1}, \\
\frac{\partial\Pi_v}{\partial\lambda_i} &= \frac{1}{\log_2N}[\log_2\lambda_i - \log_2(1-\sum\lambda_j)]\nonumber\\
&=\frac{1}{\log_2N}[\log_2\lambda_i - \log_2\lambda_N], \\
\frac{\partial\Pi_b^2}{\partial\lambda_i} &= -N^N\lambda_1...\lambda_{i-1}\lambda_{i+1}...\lambda_{N-1}(1-\sum\lambda_j - \lambda_i) \nonumber\\
&=N^N\lambda_1...\lambda_{i-1}\lambda_{i+1}...\lambda_{N-1}(\lambda_i - \lambda_N).
\end{align}
Note that $\lambda_i - \lambda_N$ will always have the same sign as $\log_2\lambda_i - \log_2\lambda_N$, since $\lambda_i > \lambda_N \iff \log_2\lambda_i > \log_2\lambda_N$. Therefore the derivatives $\frac{\partial\Pi_s}{\partial\lambda_i}$ and $\frac{\partial\Pi_v}{\partial\lambda_i}$ must have the same sign. Moreover, assuming none of the eigenvalues are zero, we see that $\frac{\partial\Pi_s}{\partial\lambda_i}$ and $\frac{\partial\Pi_b^2}{\partial\lambda_i}$ both equal a positive number multiplied by $\lambda_i - \lambda_N$, and therefore will also have the same sign. 

Therefore for any given point in the eigenvalue space, the partial derivatives of the measures $\Pi_s$, $\Pi_{sskf}$, $\Pi_v$ and $\Pi_b$ will always have the same sign, yielding the similar graphical behaviour exhibited in figure \ref{puritygraphs}. 

This implies that the purity measures above will behave similarly so long as we are \emph{varying only two eigenvalues} ($\lambda_i$ and as a consequence, $\lambda_N$). If we vary more eigenvalues simultaneously, then in general, each measure behaves differently. In other words, if we use the aforementioned measures to compare the purity of two quantum states, they will all agree which state is purer as long as the two states differ in only two eigenvalues. If the two states differ in three or more eigenvalues, the measures will, in general, not agree which is purer. This is clearly illustrated in table \ref{puritytablevalues}.

\bibliographystyle{unsrt}
\bibliography{PurityBib}

\end{document}